\documentclass[twocolumn]{aastex63}
\usepackage[utf8]{inputenc}
\usepackage{comment}
\usepackage{amsmath}
\newcommand\numberthis{\addtocounter{equation}{1}\tag{\theequation}}
\usepackage{natbib}
\setcitestyle{super ,open={[},close={]}} 

\begin{document}
\title{Divergent Reflections around the Photon Sphere of a Black Hole}

\author[0000-0002-5460-6126]{Albert Sneppen}
\affiliation{Niels Bohr Institute, University of Copenhagen, Blegdamsvej 17, K\o benhavn \O~2200, Denmark}
\affiliation{Cosmic Dawn Center (DAWN)}

\begin{abstract}
From any location outside the event horizon of a black hole there are an infinite number of trajectories for light to an observer. Each of these paths differ in the number of orbits revolved around the black hole and in their proximity to the last photon orbit. With simple numerical and a perturbed analytical solution to the null-geodesic equation of the Schwarzschild black hole we will reaffirm how each additional orbit is a factor $e^{2 \pi}$ closer to the black hole's optical edge. Consequently, the surface of the black hole and any background light will be mirrored infinitely in exponentially thinner slices around the last photon orbit. Furthermore, the introduced formalism proves how the \textit{entire} trajectories of light in the strong field limit is prescribed by a diverging and a converging exponential. 
Lastly, the existence of the exponential family is generalized to the equatorial plane of the Kerr black hole with the exponentials dependence on spin derived. Thereby, proving that the distance between subsequent images increases and decreases for respectively retrograde and prograde images. In the limit of an extremely rotating Kerr black hole no logarithmic divergence exists for prograde trajectories. \newline
\end{abstract}

\section{Introduction}
Black holes are famously objects where the spatial paths of light are drastically bent by the curvature of space-time. While light itself cannot escape the central mass at the event horizon, at further distances light may orbit the black hole. In the generic case of a non-rotating and electrically neutral black hole [ie. a Schwarzschild black hole \cite{Schwarzschild1916}] the event horizon is located at radial coordinate $R_s=\frac{2GM}{c^2}$, while photons may follow unstable circular orbits at $\frac{3}{2} R_s$, which is the so-called photon-sphere or last photon orbit. Any photon orbiting below this distance will plunge into the black hole, while light that remains further away will spiral out towards infinity.

However, depending on the photon's proximity to the last photon orbit it may complete several orbits before spiralling into the event horizon or out towards infinity \cite{Luminet1979,Stuckey1993}. As we approach the limit where the photons graze the exact critical orbital radius the photon will orbit an infinite number of times. Inversely, from the perspective of an observer at infinity this implies that light from any point (from the event horizon to the background) may orbit the black hole an arbitrary number of times. For each of these paths the light will reach the observer slightly closer to the edge of the black hole's shadow \cite{Perlick2010}. Therefore, the observer will see the entire surface of the event horizon and the entire universe repeating infinitely near the edges of the black hole. This infinite mapping has been extensively studied with the deflection angle diverging logarithmically in the strong field limit (see \cite{Darwin1959,Luminet1979,Bozza2002,Perlick2010,Munoz2014,Gralla2019}). 


However, we present a methodology, which differs from previous research by reformulating the trajectory of light in terms of a second order differential equation and quantifying its linear stability. In \S \ \ref{SimOrb} we investigate how small deviations away from the optical edge of a black hole behave with a ray-tracing algorithm. We supplement with an analytical derivation in \S \ \ref{LinSta}. In both the numerical and analytical case we will show that small perturbations grow exponentially. Inversely, each additional orbit will be mapped to an exponentially thinner ring, with each subsequent image a factor $e^{2 \pi}$ thinner. Ultimately, this paper investigates a well known problem from a new analytical perspective suggesting not only the deflection angle but the entire trajectories of light near the photon-sphere are prescribed by two duelling exponential functions. 

Crucially, this approach is generalizable to any spherically symmetric black hole or even the equatorial plane of a spinning black hole. In \S \ \ref{GenKerr}, the exponentials dependence on the spin is derived and illustrated. Here it is proved for the first time that the spatial frequency of prograde and retrograde images will respectively increase and decrease from the Schwarzschild case. 



\begin{figure*}[t!]
\begin{center}
\includegraphics[width=0.98\linewidth]{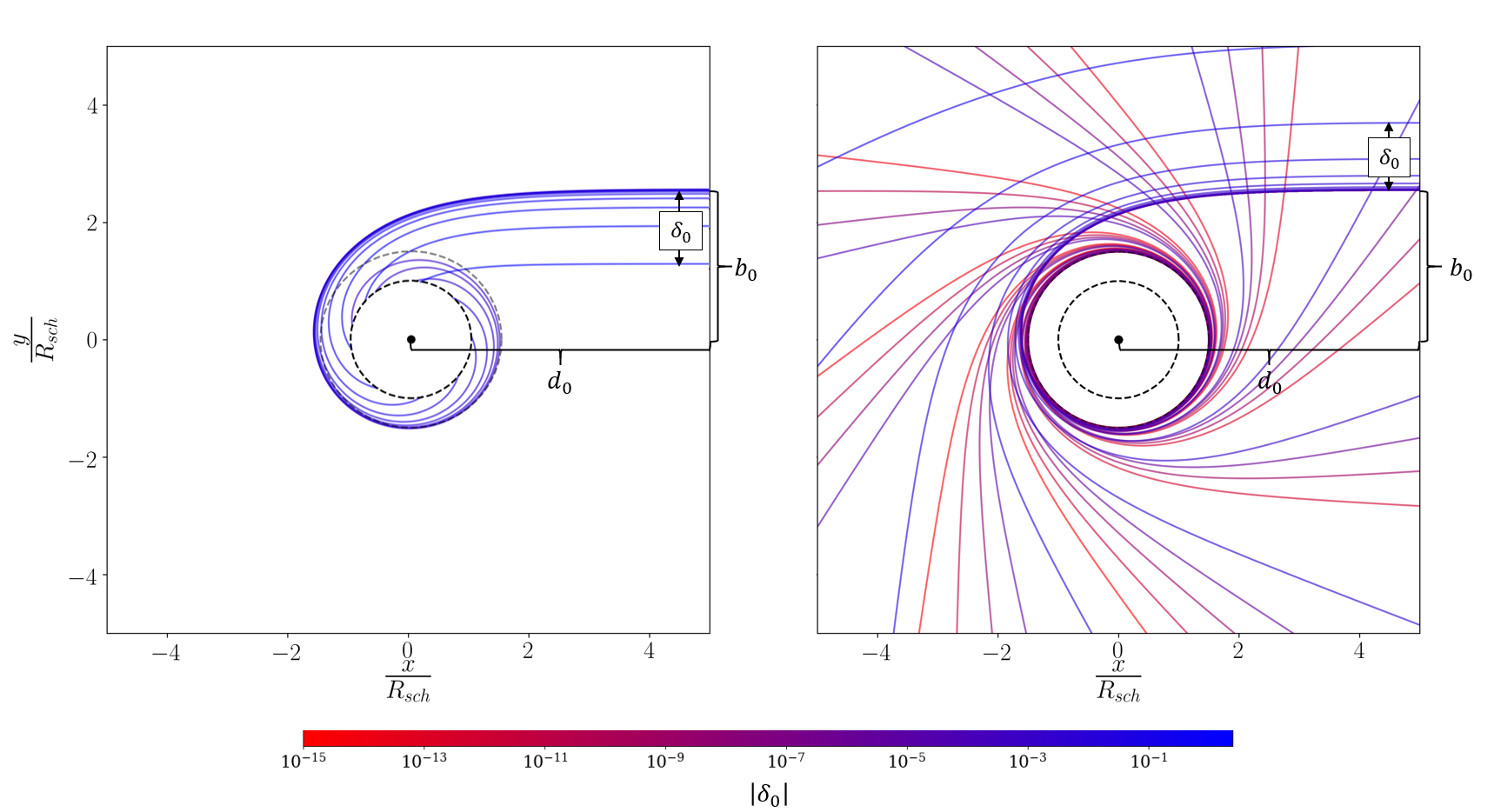}
\caption{Simulated rays of light satisfying Eq. \ref{eq:endu} with $\delta_0 < 0$ (left) and $\delta_0 > 0$ (right) with coloring indicating magnitude of $\delta_0$. The black hole is shaded in grey with the last photon orbit indicated with a dotted grey line. Each successive light-trajectory plotted is a factor of 2 closer to the photon capture radius with the resulting deflection angle increasing just below 40 degrees. Thus, the logarithmic scaling towards the photon capture radius maps to a linear evolution in $\phi$. }  
\label{raysoflight}
\end{center}
\end{figure*} 

\newpage 

\section{Analytical Setup}
The Schwarzschild metric has the form in units with the speed of light, $c=1$: 

\begin{equation}
    ds^2 = \Big( 1-\frac{R_s}{r} \Big) dt^2 - \Big( 1-\frac{R_s}{r} \Big)^{-1} dr^2 - r^2 d\theta^2 - r^2 \sin^2(\theta) d\phi^2
\end{equation}


Without loss of generality we can set the orbital plane of the light $\theta=\frac{\pi}{2}$. Introducing the 2 conserved quantities, angular momentum and energy, we can for a mass-less particle reduce the equation for the trajectory of light to (\cite{Musiri2019,Stuckey1993}):

\begin{equation}
    \Big( \frac{dr}{d\phi} \Big)^2 = -\Big( 1-\frac{R_s}{r} \Big) r^2 + \frac{r^4}{b^2}
    \label{metric}
\end{equation}

Here $b$ is the constant ratio of a photon's angular momentum to energy. Rewriting the differential with $\phi$ is only applicable if there is an angular evolution, so Eq. \ref{metric} does not apply in the limiting and trivial case of light moving radially towards or away from a black hole. Using the substitution $u = R_s/r$ and differentiating both sides with $\frac{d}{d\phi}$ yields the simple second order equation: 
\begin{equation}
    \frac{d^2u}{d\phi^2} = \frac{3}{2} u^2 - u
    \label{eq:endu}
\end{equation}

We can immediately reproduce the stationary orbit of the photon sphere for the Schwarzschild black hole by setting $\frac{d^2u_{eq}}{d\phi^2} = 0$: $r_{eq} = \frac{R_s}{u_{eq}} = \frac{3}{2} R_s$. Note, the trivial equilibrium solution for $u=0$ (ie. at infinite distances from the black hole) will not be discussed further.

\section{Simulated Orbits}
\label{SimOrb}

Given the differential equation (eq. \ref{eq:endu}) relating distance to the angular deflection we can numerically integrate Eq. \ref{eq:endu} using quartic Runga-Kutta (see Fig. \ref{raysoflight}). In this approach we are propagating the light from an observer to the black hole or the background universe. This yields the same light-path as the opposite direction, because the solutions of Eq. \ref{metric} are independent of the direction of the light. Integrating Eq. \ref{eq:endu} requires 2 initial conditions: In Cartesian coordinates we center the black hole at the origin, set the initial direction of light to be $\hat{v}_0 = (-1, 0)$ and the initial position $\vec{r_0} = (d_0, b_0+\delta_0)$. Here $b_0$ (the critical impact parameter) is the distance within which photons are captured and outside which photons are deflected. Here $d_0$ can be arbitrarily large given $b_0$ at that distance which in the $\lim_{d_0 \to \infty} b_0 = \frac{\sqrt{27}}{2} R_s$ becomes the photon capture radius commonly found in literature \cite{Schutz2009}. $\delta_0$ is our initial perturbation which we use to avoid the ambiguity of defining the closest approach for light rays that spiral within the photon sphere. 
Importantly, $\delta_0$ is interpretable as how far the observer is looking away from rim of the optical black hole.

The path for light with varying $\delta_0$ can be seen in Fig. \ref{raysoflight}. Positive and negative perturbations will respectively spiral out to infinity or plunge into the event horizon as expected. As $\delta_0$ becomes smaller the deflection angle increases. Note, $\phi$ increases linearly when moving logarithmically closer to the photon capture radius.  

\begin{figure}
\begin{center}
\includegraphics[width=\linewidth]{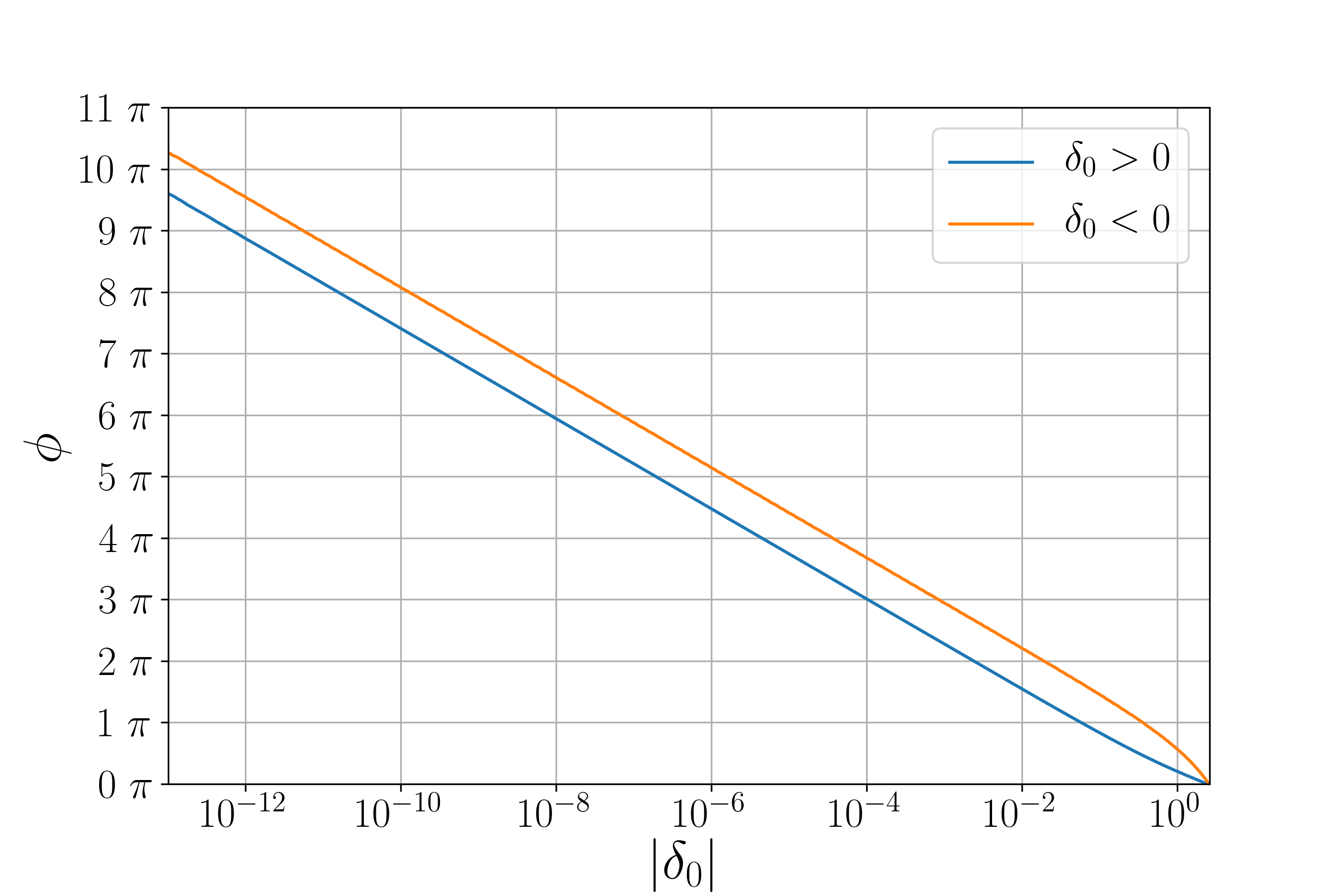}
\caption{Angle of rotation for simulated light-rays as a function of deviations from the equilibrium (in dimensionless units with $\delta_0 = l/R_s$) with $\phi=0$ representing unbent light-rays. For both small positive and negative perturbations a clear exponential relation to $\phi$ is visible. }  
\label{Deviations}
\end{center}
\end{figure} 

In Fig. \ref{Deviations} the results can be seen for positive (where the angle $\phi$ is the unwrapped deflection angle) and negative perturbations (with the angle $\phi$ being defined as the angle orbited around the black hole at the time when photons cross the event horizon). For large perturbations ($|\delta_0|>10^{-2}$) 
the relationship between angle and distance is not simply exponential. However in the small perturbation regime ($|\delta_0| < 10^{-2}$) a tight exponential relationship is visible. To determine the exponent in the exponential regime, we can fit, $\phi = s \ln(\delta_0) + c$, and with slope: $s = -1.0000 \pm 0.0001 $. 
Inverting this expression for $\delta_0$ implies that to achieve another orbit requires being a factor of $f = e^{-2 \pi s} = 535.60 \pm 0.45$ closer to the optical edge of the black hole.

\begin{figure*}[t!]
\begin{center}
\includegraphics[width=\linewidth]{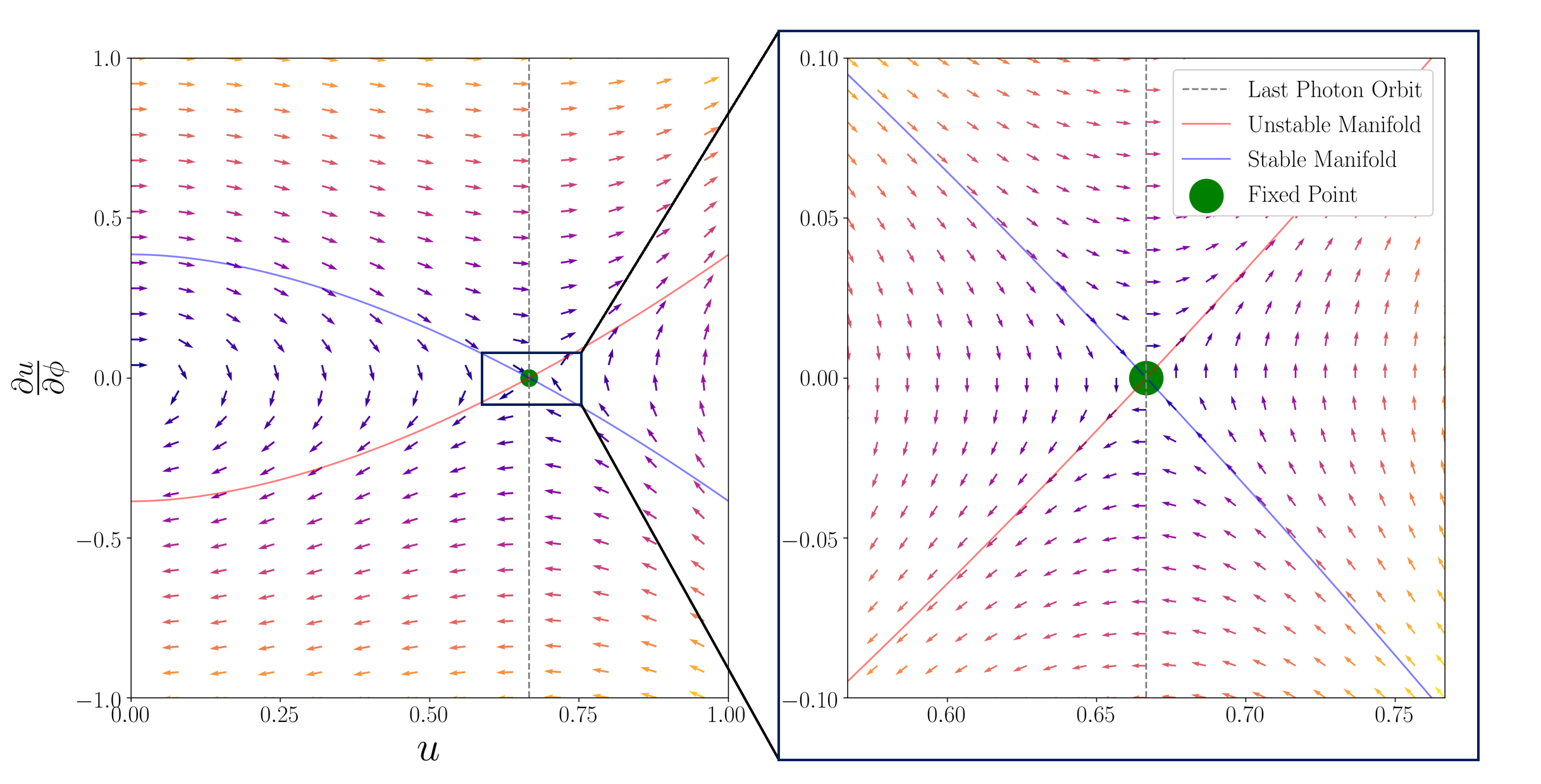}
\caption{ Entire (left) and zoomed-in (right) phase-space portrait for light trajectories obeying Eq. \ref{eq:endu} with the arrows' coloring indicating the magnitude of change (brighter hues implies longer vectors). $u = 1$ is the event-horizon, $u=0$ represents infinity and $u=\frac{2}{3}$ is at the photon sphere. 
If $u=\frac{2}{3}$ and $\frac{d u}{d \phi}=0$ the photons are on circular orbit, so this represents a fixed point. Notable, this is not a stable fixed point as deviations will in general grow. The stable and unstable manifolds are drawn which in the enlarged version are approximately linear. The stable manifold evidently represents a separatrix between the initial conditions of trajectories which will cross the event horizon or be ejected to infinity. Thus, the stable manifold is equivalent to the optical rim of the black hole. } 
\label{phasespace}
\end{center}
\end{figure*} 

\section{Linear Stability}
\label{LinSta}

To interpret these numerical results we will utilise linear stability analysis by adding small perturbations, $u \rightarrow u_{eq} + \delta$ to the equilibrium solution of Eq. \ref{eq:endu}.

\begin{equation}
    \frac{d^2(u_{eq}+\delta)}{d\phi^2} = \frac{3}{2} (u_{eq}+\delta)^2 - (u_{eq}+\delta)
    \label{eq:perturb1}
\end{equation}

Linearlizing the equation around $u_{eq} = \frac{2}{3}$ one gets: 

\begin{equation}
    \frac{d^2\delta}{d\phi^2} = \delta + \frac{3}{2} \delta^2 \approx \delta
    \label{eq:perturb}
\end{equation}

Which has the solution: 

\begin{equation}
\delta = \delta_1 e^{\phi} + \delta_{-1} e^{-\phi}  
\label{eq:perturbsol_sch}
\end{equation}

Evidently the first term grows in magnitude while the latter decreases, with the constants $\delta_1$ and $\delta_{-1}$ determining in which regime each term dominates. The constants are set by the initial conditions of the trajectory, which will be discussed further in \S \ 4.2. Note the dual exponential form is to be expected as the equilibrium solution is a saddle point.  

\subsection{Intuition through manifolds}
An alternate perspective on these exponential solutions is in the phase-space of Eq. \ref{eq:endu}. This is shown in Fig. \ref{phasespace} where for every initial condition ($u,\frac{du}{d\phi}$) a vector is plotted indicating the angular change in both variables (ie. $\frac{d u}{d \phi},\frac{d^2 u}{d \phi^2}$). The trajectories terminating at $u=1$ (ie. $r=R_s$) are the rays of light reaching the event horizon, while infinity is at $u=0$. Most trajectories will cross the photon-sphere with radial velocities, but if $\frac{d u}{d \phi}=0$ on the photon-sphere then the photon will stay in its circular orbit indefinitely. Thus, orbits on the photon-sphere represent a fixed point in the phase-space.  

The set of initial conditions which converge towards the photon sphere (which is called the stable manifold) is indicated with a blue line. Photons on this trajectory will asymptotically approach the photon sphere. Conversely, the unstable manifold (ie. the set of initial conditions which reach the fixed point for $\phi \to - \infty$) is plotted in red. The symmetry between stable and unstable manifolds seen in the phase-space is due to the Schwarzschild metric and therefore Eq. \ref{eq:perturbsol} being independent of the direction of time.

As seen in Fig. \ref{phasespace} (left) for the stable and unstable manifolds $\frac{\partial u}{\partial \phi}$ is in general not linear in $u$, but when we are close to the fixed point ($\delta^2 < |\delta|$, see Eq. \ref{eq:perturb}), the relationship becomes approximately linear. Importantly, there are two sets of eigenvectors around the photon sphere. The first with an eigenvalue of $-1$ (the exponentially approaching term) and the unstable manifold with an eigenvalue of $+1$ (the exponentially diverging term). Thus, the phase space clearly follows the intuition of Eq. \ref{eq:perturbsol_sch}. 

The different signs of the eigenvalues proves that the fixed point is a saddle-point. A saddle-point is inherently unstable as a perturbation from the photon sphere will generically result in an exponential divergence. Evidently, the positive eigenvalue implies that a trajectory will diverge exponentially from the bound orbit with a factor $e^{\pi n} = e^{\gamma n}$ for each half orbit n. Here the Lyapunov exponent, $\gamma$ [defined in \cite{Johnson2020}] characterizes the instability of the bound orbit relative to a half-orbit n. Thus, for the Schwarschild case for the photon sphere, $\gamma=\pi$. 

Lastly, notice the eigenvalues around the fixed point in $u$ are also the eigenvalues for $r$ as $\frac{\partial u}{\partial \phi} = u \Rightarrow \frac{\partial r}{\partial \phi} = -r$. Therefore, the $\pm 1$ eigenvalues in ($u,\frac{\partial u}{\partial \phi}$) corresponds to the eigenvalues $\mp 1$ in ($r,\frac{\partial r}{\partial \phi}$).   

\subsection{A tail of two exponentials}
\begin{figure}
\begin{center}
\includegraphics[width=\linewidth]{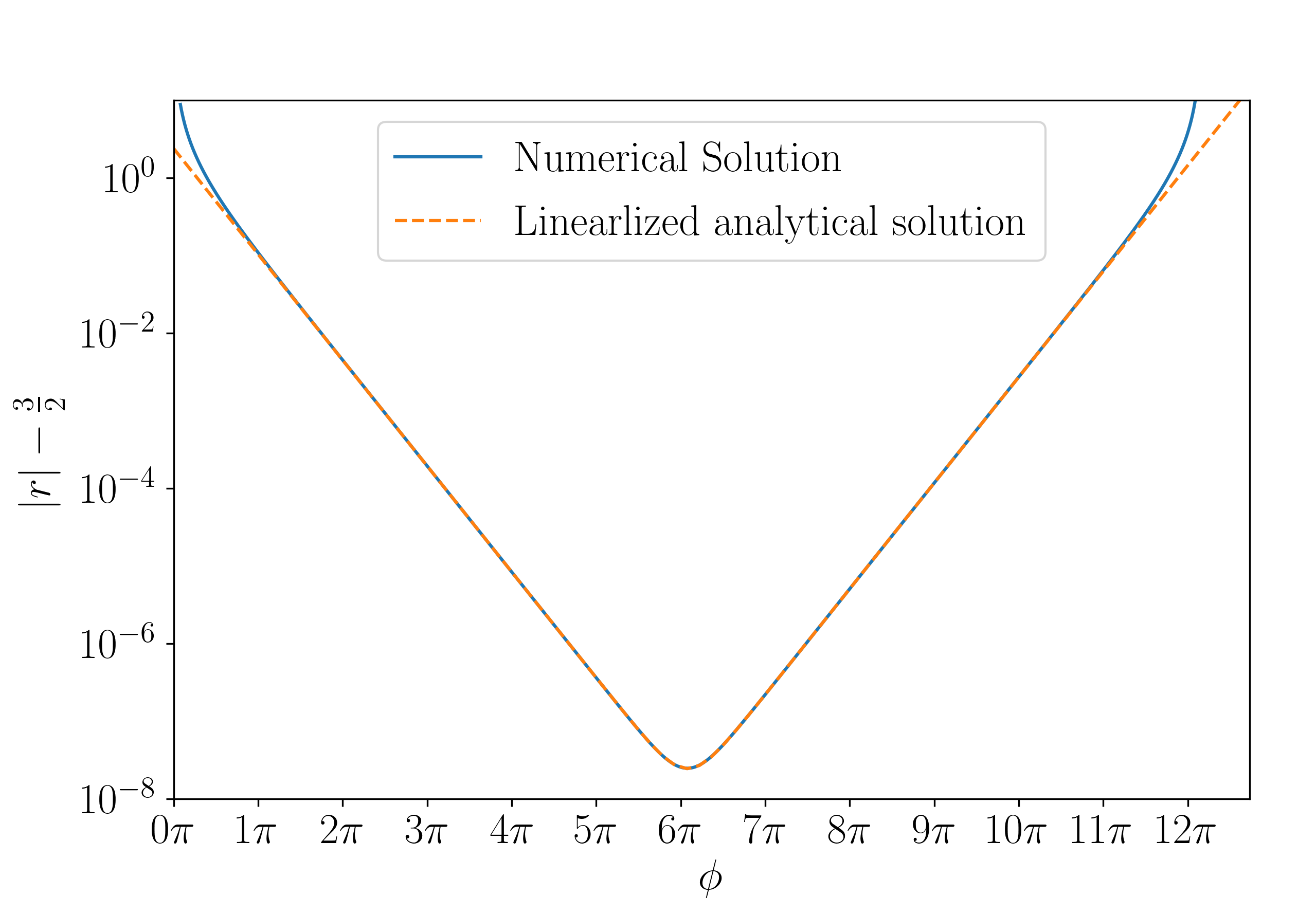}
\caption{ The radial distance between a light-ray (with $\delta_0 = 10^{-15}$) and the last photon orbit as a function of deflection angle [in blue]. The predicted analytical combination of an exponentially declining ($e^{-\phi}$) and exponentially growing ($e^{\phi}$) term is indicated with a yellow dashed line. Evidently, each term dominates at different angles of $\phi$, with fitted lines suggesting $\delta_1 \approx 10^{-16}$ and $\delta_{-1} \approx 1$. For $u-u_{eq} \approx 1$, the linearised solution no longer holds. }  
\label{Distance}
\end{center}
\end{figure}

Given Eq. \ref{eq:perturbsol_sch} we find the linearlized solutions: 

\begin{equation}
u = u_{eq} + \delta_1 e^{\phi} + \delta_{-1} e^{-\phi}  
\label{7}
\end{equation}

\begin{equation}
\frac{du}{d \phi } = \delta_1 e^{\phi} - \delta_{-1} e^{-\phi}  
\label{8}
\end{equation}
When investigating the trajectory of light close to the black hole both exponential terms are needed to cross the equilibrium distance (see Eq. \ref{7}) or for $\frac{d u}{d \phi }$ to change sign (as seen in Eq. \ref{8}). The importance of both exponential terms is also illustrated in Fig. \ref{Distance}, where the light approaches the photon sphere exponentially (with each rotation bringing it a factor $e^{2 \pi}$ closer) until at a crossover-angle of $\phi_c \approx 6 \pi$. After this the divergent $e^{\phi}$ dominates and the light is ejected towards infinity. If $\delta_{1}$ had the opposite sign then $\frac{du}{d\phi}$ would remain negative so the crossover-angle would be on the last photon orbit, after which the light would diverge exponentially from the photon-sphere towards the black hole. Curiously, this implies the angle swept by the ray around the black hole prior to the photon sphere is similar to the angle swept by the ray from the photon-sphere to the event horizon.

Notably, the light-ray on the trajectory exactly on the rim of the black hole's shadow (ie. $\delta_0=0$) is the solution which is exponentially approaching the photon-sphere indefinitely as it neither diverges towards the black hole or the background universe. It follow that the convergent exponential and in extension $\delta_{-1}$ must be independent of $\delta_{0}$. Instead $\delta_{-1}$ is set by the approximate distance where the linearised expression holds ($\delta_{-1} \approx 1$). Any deviations from the critical impact parameter, $\delta_0 \neq 0$, will grow exponentially, which implies $\delta_1$ (the divergent exponential) is set by $\delta_0$. Thus, the order of magnitude estimates neatly follow the fitted lines in Fig. \ref{Distance}.


While the derivation is only applicable in the linearized regime the implications reach beyond the immediate surroundings of the photon-sphere, as the total deflection of light may be dominated by the angular rotation, while the photons are in the linearized regime. When investigating the total deflection angle or angle of rotation for light (as seen \S \ 3) we are solving the trajectories for light moving away from $u_{eq}$, where the divergent exponential must dominate. Each additional orbit of light will be mapped a factor $f = e^{2 \pi}$ nearer the rim of the black hole's shadow, because decreasing $\delta_0$ by a factor $e^{2\pi}$ delays the exponentially growing term exactly one orbit. Furthermore, it should be noted that the predicted analytical value, $f = e^{2 \pi} \approx 535.49$, is remarkably close to the numerically fitted relationship seen in \S \ \ref{SimOrb}. 

\begin{figure*}
\includegraphics[width=0.99\linewidth]{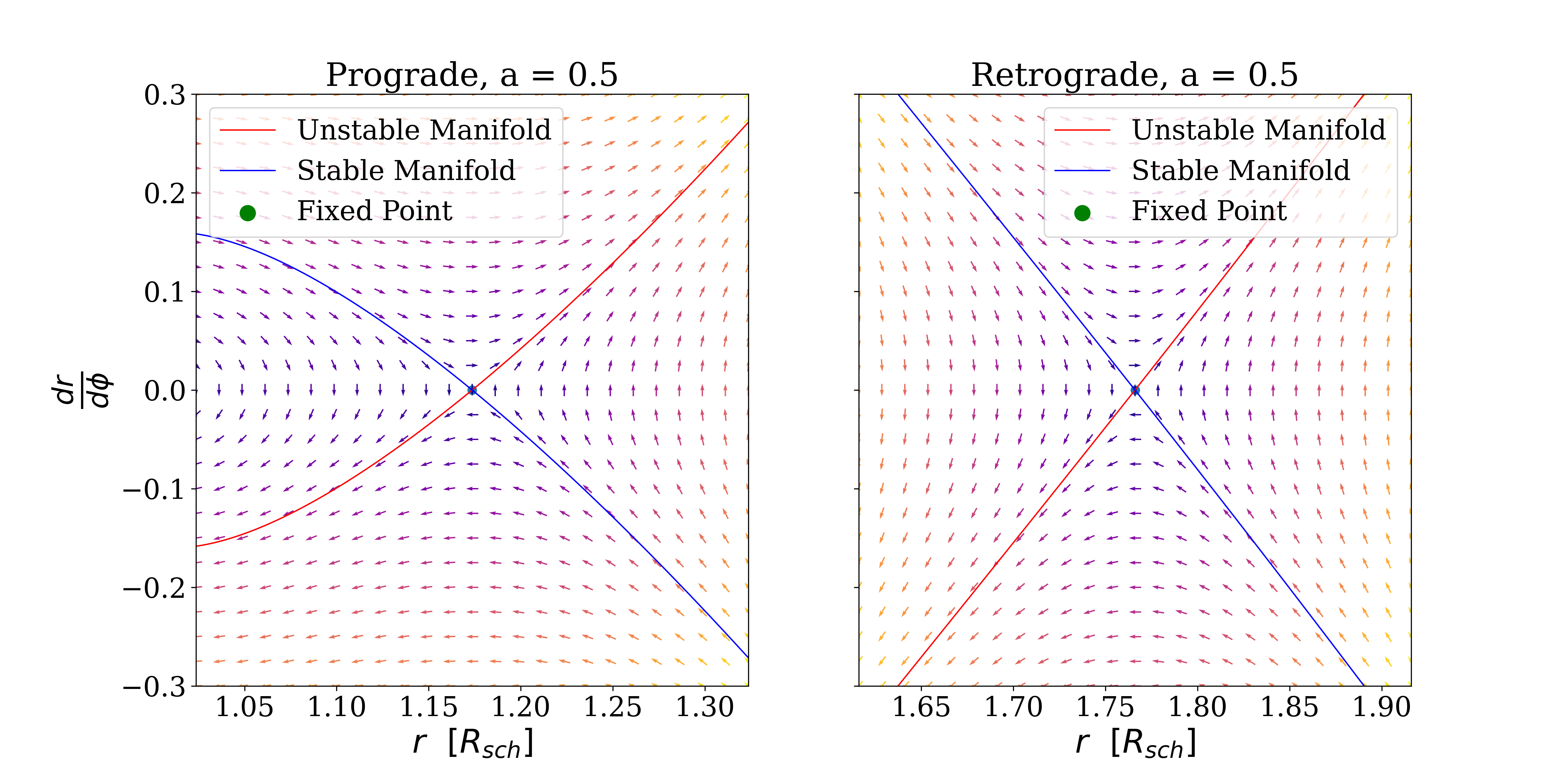}
\caption{Phase-space portrait for light trajectories obeying Eq. \ref{eqn:kerr} with a=0.5 (left: prograde and right: retrograde) with the arrows' coloring indicating the magnitude of change (brighter hues imply longer vectors). The stable and unstable manifolds are drawn which behave approximately linear with a flatter and steeper slope than Fig. \ref{phasespace} for respectively the prograde and retrograde orbits. Note the substitution $u=\frac{1}{r}$ does not remove the critical impact parameter unlike the Schwarzschild case. Therefore the figure remains in $r$ not $u$ like Fig. \ref{phasespace}. }  
\label{fig:Kerr-phaseportrait}
\end{figure*} 

Additionally, for deflected light it is noteworthy that the closest approach to the photon-sphere will only decrease by a factor of $e^{\pi}$ for each additional orbit, because the cross-over angle is set by the intersection of the two exponential terms. Similarly, for light crossing the photon-sphere the angle swept from the event horizon to the photon-sphere is similar to the angle swept from the photon-sphere to the observer, as the cross-over angle is still defined by the intersection.   \vspace{15mm}

\section{Generalization to Kerr Metric}
\label{GenKerr}
It should be emphasised, that the Schwarzschild metric is the limiting case of a non-spinning black hole. Without this requirement one gets the so-called Kerr metric (Here written in Boyer–Lindquist coordinates with $\Sigma=r^2+a^2 cos(\theta)$ and $\Delta = r^2 - R_s r + a^2$):  

\begin{align*} 
ds^2 &= \Big( 1-\frac{R_s r}{\Sigma} \Big) dt^2 + \frac{2 a R_s r}{\Sigma} sin(\theta)^2 dt d\phi \\
    &- \frac{(r^2+a^2)^2 - \Delta a^2 sin(\theta)^2}{\Sigma} sin(\theta)^2 d\phi^2 - \frac{\Sigma}{\Delta} dr^2 - \Sigma d\theta^2
\end{align*}

Here $0 \leq a \leq 1$ is the angular momentum factor, so naturally the Kerr metric reduces to the Schwarzschild metric for $a=0$. For orbits in the equatorial plane (where a 2-dimensional analysis is still an exhaustive description) we set $\theta=\frac{\pi}{2}$. Further deliberation on non-equatorial orbits may be found through elliptic integrals as expressed in \cite{Gralla2020}. Introducing the 2 conserved quantities, angular momentum and energy, the trajectory of photons reduces to (\cite{Cramer1997}):

\begin{equation}
    \Big( \frac{dr}{d\phi} \Big)^2 = \frac{( r^2 - R_s r + a^2)^2 \left[ 1 + \frac{R_s}{r^3} (b-a)^2 - \frac{1}{r^2} (b^2 - a^2)  \right] } {\left[ \frac{R_s a}{r} + (1-\frac{R_s}{r})b \right]^2}
\end{equation}

With $b$ once more being the constant ratio of a photon's angular momentum to energy. Differentiating both sides with $\frac{d}{d\phi}$ yields a second order differential equation. 

\begin{align*} 
    \frac{d^2r}{d\phi^2} &=  \frac{(2r-R_s)(r^2 - R_s r + a^2) \left(1 + \frac{R_s(b-a)^2}{r^3} - \frac{b^2 - a^2}{r^2} \right)}{\left[ \frac{a R_s}{r} + (1-\frac{R_s}{r})b \right]^2} \\
    &+ \frac{\frac{a R_s-b R_s}{r^2}(r^2 - R_s r + a^2)^2 \left(1 + \frac{R_s(b-a)^2}{r^3} - \frac{b^2 - a^2}{r^2} \right)}{\left[ \frac{a R_s}{r} + (1-\frac{R_s}{r})b \right]^3} \\
    &+ \frac{(r^2 - R_s r + a^2)^2 \left(1 + \frac{3 R_s(b-a)^2}{r^4} - \frac{2(b^2-a^2)}{r^3} \right)}{2 \left[ \frac{a R_s}{r} + (1-\frac{R_s}{r})b \right]^2} \numberthis \label{eqn:kerr}
\end{align*}

The phase portrait for Eq. \ref{eqn:kerr} for $a=0.5$ is illustrated in Fig. \ref{fig:Kerr-phaseportrait}. As before, the fixed point is set by the roots of $\frac{dr^2}{d\phi^2}$, which will depend on $a$. In contrast to the Schwarzschild parameterization, Eq. \ref{eqn:kerr} depends on $b$, so the critical impact parameter of prograde and retrograde orbits must be evaluated at any $a$ (for clarification see \cite{Cramer1997}).

\begin{equation}
    b_{retrograde} = - 3 R_s \cos \left(\frac{\arccos(a)}{3} \right) - a
    \label{eq_bret}
\end{equation}
\begin{equation}
    b_{prograde} = 3 R_s \cos \left(\frac{\arccos(a)}{3} \right) - a
    \label{eq_bpro}
\end{equation}

Retrograde and prograde orbits are obtained by evaluating the critical impact parameter with Eq. \ref{eq_bret} and \ref{eq_bpro}. For any given spin we may then determine the fixed points, $r_{eq}$ such that $\frac{\partial r^2}{\partial  \phi^2}$=0. One root, $\Delta=0$, representing the event horizon with the remaining real root describing the photon circle (see Fig. \ref{fig:s}). Linearizing generically yields: 

\begin{figure}[t!]
\begin{center}
\includegraphics[width=0.99\linewidth]{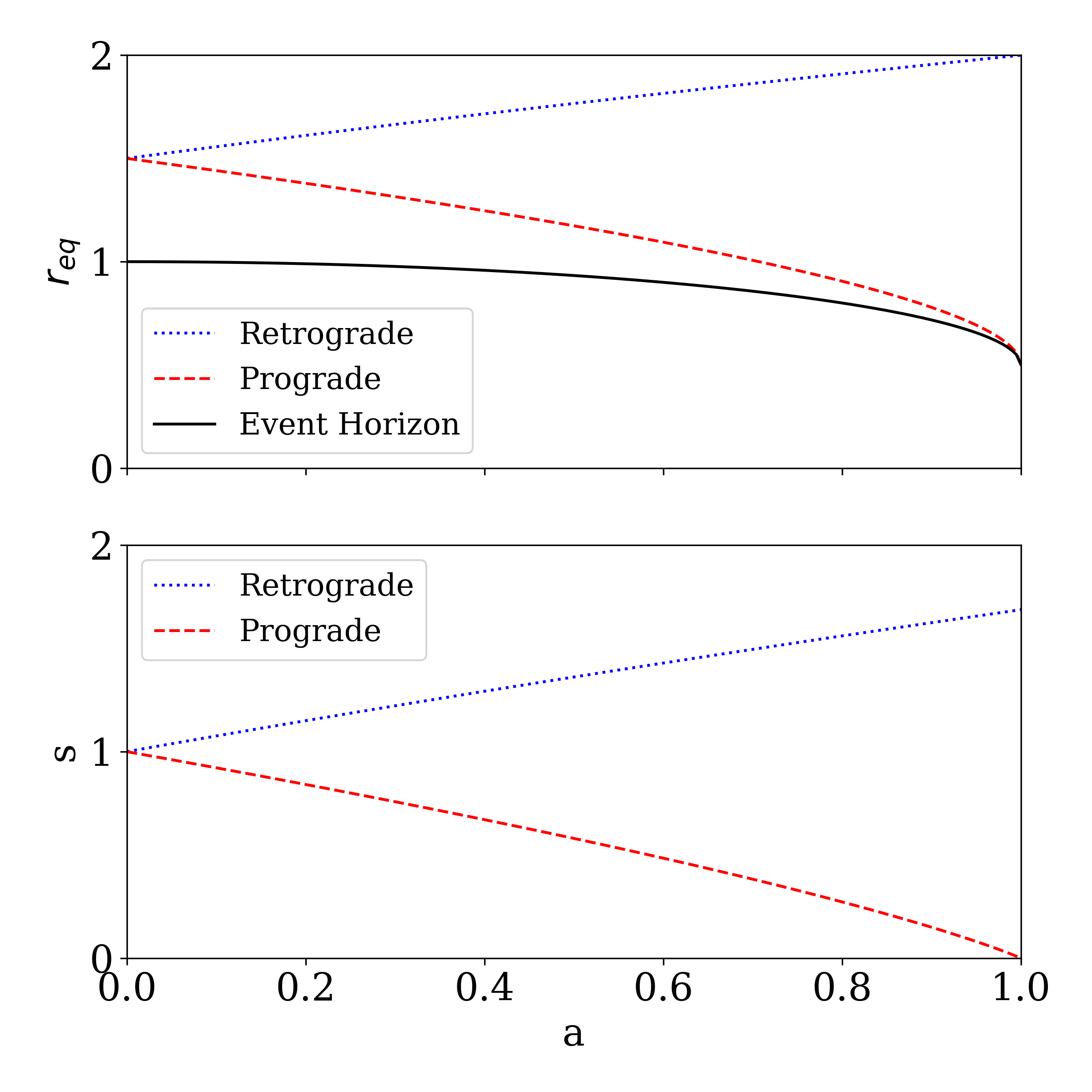}
\caption{Radii of photon circle, $r_{eq}$ (top), and linear coefficient of Taylor-expansion, $s$ (bottom). Retrograde orbits (in dotted blue) and prograde orbits (in dashed red).  The location of the photon circle is in agreement with \cite{Bardeen1972}, while the generic result $s>0$ for $0 \leq a < 1$, implies that the fixed point will always be a saddle point with eigenvalues $\pm \sqrt{s}$ and therefore unstable. Notable at $a=1$ the two real fixed points of prograde motion bifurcate}  
\label{fig:s}
\end{center}
\end{figure} 

\begin{equation*}
    \frac{dr^2}{d\phi^2} =  s \cdot (r - r_{eq}) + O( (r-r_{eq})^2 )
\end{equation*}

Importantly, we once more we get a family of two exponentials and for the first time derive the exponential unwinding for the strong field limit of light:

\begin{equation}
r = r_{eq} + \delta_1 e^{\sqrt{s} \phi} + \delta_{-1} e^{- \sqrt{s} \phi}  
\label{eq:perturbsol}
\end{equation}

For any spin, $a$, the exponential coefficient, $s$ is shown in Fig. \ref{fig:s}. Evidently the fixed point is still a saddle point and therefore unstable. Here, the exponential divergent term corresponds to Lyapunov exponent of $\gamma = \pi \sqrt{s}$ as trajectories will diverge from the photon sphere with a factor of $e^{\pi \sqrt{s}}$ over a half orbit. Notable, the exponential coefficient $s$ results in a even faster divergence of the logarithmic angle for retrograde orbits. For $\lim_{a \rightarrow 1}(s_{retrograde})=\frac{27}{16}$, so another orbit would require being a factor of $f = e^{2 \pi \sqrt{\frac{27}{16}}} \approx 3500$ closer to the optical edge of the black hole. Conversely, for prograde orbits the exponential function unwinds evermore slowly for larger spins. 
For a black hole spinning with $a=0.99$ (as potentially observed \cite{McClintock2006}), where $s=0.012$, only a factor $f = e^{2 \pi \sqrt{0.012}} \approx 2$ is required. Here, each repeated image would merely be a factor of 2 closer to the optical edge of the black hole. In the limit of an extreme Kerr field, $\lim_{a \rightarrow 1}(s_{prograde}) = 0$, the eigenvectors collapse and the fixed point becomes a degenerate node. Thus, an extremely rotating Kerr black hole has no exponential trajectories for the prograde motion. 

Thus, when viewing the equatorial plane of a spinning black hole both prograde and retrograde reflections display the exponential repetition, but prograde copies of a source will repeat rapidly compared to the retrograde copies. This asymmetry has potentially far-reaching applications to observables as any observational signature is limited by the brightness of subsequent images decreasing sharply (see \cite{Johnson2020}). Therefore, the rapid spatial repetition of prograde images will provide the first observational signatures of the exponential repetition within detection capabilities. 

Lastly, the mathematical generality of two real eigenvalues existing for all $a$ should not go unstated. Regardless of the spin of the black hole, there will always exist a family of a convergent and divergent exponential. These exponentials prescribe the entire trajectories of light near the photon orbits. Their prescription implies that any source object in the plane be repeated in an exponentially thinner series of copies, with the scale of repetitions set by the spin of the black hole. \newline

\section{Conclusion}
\label{Conclusion}
This work introduces a family of two distinct exponential solutions which together provide a succinct description of the \emph{entire} orbital trajectories of light near a Schwarzschild black hole. Thereby we provide analytical insight into the solutions previously developed (see \cite{Darwin1959,Luminet1979,Bozza2002,Perlick2010,Munoz2014,Gralla2019}). 

Our formalism provides a few important interpretations. Firstly, it states that the deflection angle of background light will diverge logarithmically when the trajectory approaches the last photon orbit. Equivalently, from the perspective of a distant observer looking at the optical edge of the black hole (the photon capture radius) the entire background will be mapped to exponentially thinner rings. Secondly, the event horizon of the black hole itself will be mapped repeatedly in exponentially thinner rings just inside the photon capture radius. Therefore, any object accretting onto the black hole may be observed repeatedly nearer and nearer the optical edge. Thirdly, this edge of the black hole is the location of both the stable and unstable manifold. 

The proof presented here is immediately generalizable to any spherically symmetric space-time (such as a Reissner-Nordstrøm black hole). Such metrics can similarly be written as a second order differential equation in $r$ with steady state and perturbed solutions. Further work may investigate these exponentials, which will in general be characterised by a constant, $s \neq 1$, to be multiplied on $\phi$ in the exponents of Eq. \ref{eq:perturbsol}. 

Importantly, as seen in \S \ \ref{GenKerr}, our methodology may even be applied to non-spherically symmetric black holes, such as the spinning black holes of the Kerr Metric. With increasing spin, the exponential coefficient, $s$, of prograde trajectories decreases while retrograde conversely increase. Thus, proving that the side of the black hole which rotates towards the observer repeatedly mirrors the universe in wide bands. In the limit of an extremely rotating Kerr Hole, the $s_{retrograde}=\frac{27}{16}$ and $s_{prograde}=0$. Thus, there is no logarithmic divergence for prograde reflections when $a=1$, but given any spin $a<1$, there exists an exponential family prescribing the trajectories.



Philosophically, there is a mathematical beauty within the dual exponentials of Eq. \ref{7} and Eq. \ref{eq:perturbsol}. The exponentials prescribes, that an observer at infinity will see the entire black hole's event horizon and anything accreting onto the black hole mapped infinitely when looking closer towards the photon capture radius of the black hole. Just beyond the photon capture radius, the exponentials dictate, that the observer will also see the entire universe mirrored in exponentially smaller slivers until the quantum limit. A divergence which certainly merits further reflection. \newline

The author would like to thank Mogens Høgh Jensen, Martin Pessah, Charles Steinhardt and Nikki Arendse for useful deliberations and insightful feedback. The Cosmic Dawn Center (DAWN) is funded by the Danish National Research Foundation under grant No. 140.  

\newpage
\bibliographystyle{mnras}
\bibliography{refs.bib}

\end{document}